\documentclass[12pt]{article}
\usepackage{colordvi}
\usepackage{epsfig}
\usepackage{amsmath}
\usepackage{hhline}
\usepackage{amssymb}
\usepackage{times}
\usepackage{cite}
\usepackage{helvet}

\newlength{\dinwidth}
\newlength{\dinmargin}
\begin{document}  
\newcommand{\pom}{{I\!\!P}}
\newcommand{\reg}{{I\!\!R}}
\newcommand{\slowpi}{\pi_{\mathit{slow}}}
\newcommand{\fiidiii}{F_2^{D(3)}}
\newcommand{\fiidiiiarg}{\fiidiii\,(\beta,\,Q^2,\,x)}
\newcommand{\n}{1.19\pm 0.06 (stat.) \pm0.07 (syst.)}
\newcommand{\nz}{1.30\pm 0.08 (stat.)^{+0.08}_{-0.14} (syst.)}
\newcommand{\fiidiiiful}{F_2^{D(4)}\,(\beta,\,Q^2,\,x,\,t)}
\newcommand{\fiipom}{\tilde F_2^D}
\newcommand{\ALPHA}{1.10\pm0.03 (stat.) \pm0.04 (syst.)}
\newcommand{\ALPHAZ}{1.15\pm0.04 (stat.)^{+0.04}_{-0.07} (syst.)}
\newcommand{\fiipomarg}{\fiipom\,(\beta,\,Q^2)}
\newcommand{\pomflux}{f_{\pom / p}}
\newcommand{\nxpom}{1.19\pm 0.06 (stat.) \pm0.07 (syst.)}
\newcommand {\gapprox}
   {\raisebox{-0.7ex}{$\stackrel {\textstyle>}{\sim}$}}
\newcommand {\lapprox}
   {\raisebox{-0.7ex}{$\stackrel {\textstyle<}{\sim}$}}
\def\gsim{\,\lower.25ex\hbox{$\scriptstyle\sim$}\kern-1.30ex%
\raise 0.55ex\hbox{$\scriptstyle >$}\,}
\def\lsim{\,\lower.25ex\hbox{$\scriptstyle\sim$}\kern-1.30ex%
\raise 0.55ex\hbox{$\scriptstyle <$}\,}
\newcommand{\pomfluxarg}{f_{\pom / p}\,(x_\pom)}
\newcommand{\dsf}{\mbox{$F_2^{D(3)}$}}
\newcommand{\dsfva}{\mbox{$F_2^{D(3)}(\beta,Q^2,x_{I\!\!P})$}}
\newcommand{\dsfvb}{\mbox{$F_2^{D(3)}(\beta,Q^2,x)$}}
\newcommand{\dsfpom}{$F_2^{I\!\!P}$}
\newcommand{\gap}{\stackrel{>}{\sim}}
\newcommand{\lap}{\stackrel{<}{\sim}}
\newcommand{\fem}{$F_2^{em}$}
\newcommand{\tsnmp}{$\tilde{\sigma}_{NC}(e^{\mp})$}
\newcommand{\tsnm}{$\tilde{\sigma}_{NC}(e^-)$}
\newcommand{\tsnp}{$\tilde{\sigma}_{NC}(e^+)$}
\newcommand{\st}{$\star$}
\newcommand{\sst}{$\star \star$}
\newcommand{\ssst}{$\star \star \star$}
\newcommand{\sssst}{$\star \star \star \star$}
\newcommand{\tw}{\theta_W}
\newcommand{\sw}{\sin{\theta_W}}
\newcommand{\cw}{\cos{\theta_W}}
\newcommand{\sww}{\sin^2{\theta_W}}
\newcommand{\cww}{\cos^2{\theta_W}}
\newcommand{\trm}{m_{\perp}}
\newcommand{\trp}{p_{\perp}}
\newcommand{\trmm}{m_{\perp}^2}
\newcommand{\trpp}{p_{\perp}^2}
\newcommand{\alp}{\alpha_s}

\newcommand{\alps}{\alpha_s}
\newcommand{\sqrts}{$\sqrt{s}$}
\newcommand{\LO}{$O(\alpha_s^0)$}
\newcommand{\Oa}{$O(\alpha_s)$}
\newcommand{\Oaa}{$O(\alpha_s^2)$}
\newcommand{\pt}{p_{_{\rm T}}}
\newcommand{\JPSI}{J/\psi}
\newcommand{\sh}{\hat{s}}
\newcommand{\uh}{\hat{u}}
\newcommand{\MP}{m_{J/\psi}}
\newcommand{\PO}{I\!\!P}
\newcommand{\xbj}{x}
\newcommand{\xpom}{x_{\PO}}
\newcommand{\ttbs}{\char'134}
\newcommand{\xpomlo}{3\times10^{-4}}  
\newcommand{\xpomup}{0.05}  
\newcommand{\dgr}{^\circ}
\newcommand{\pbarnt}{\,\mbox{{\rm pb$^{-1}$}}}
\newcommand{\gev}{\,\mbox{GeV}}
\newcommand{\WBoson}{\mbox{$W$}}
\newcommand{\fbarn}{\,\mbox{{\rm fb}}}
\newcommand{\fbarnt}{\,\mbox{{\rm fb$^{-1}$}}}
\newcommand{\dedx}{{\rm d} E / {\rm d} x}
%
%
\newcommand{\qsq}{\ensuremath{Q^2} }
\newcommand{\gevsq}{\ensuremath{\mathrm{GeV}^2} }
\newcommand{\et}{\ensuremath{E_t^*} }
\newcommand{\rap}{\ensuremath{\eta^*} }
\newcommand{\gp}{\ensuremath{\gamma^*}p }
\newcommand{\dsiget}{\ensuremath{{\rm d}\sigma_{ep}/{\rm d}E_t^*} }
\newcommand{\dsigrap}{\ensuremath{{\rm d}\sigma_{ep}/{\rm d}\eta^*} }
\def\Journal#1#2#3#4{{#1} {\bf #2} (#3) #4}
\def\NCA{\em Nuovo Cimento}
\def\NIM{\em Nucl. Instrum. Methods}
\def\NIMA{{\em Nucl. Instrum. Methods} {\bf A}}
\def\NPB{{\em Nucl. Phys.}   {\bf B}}
\def\PLB{{\em Phys. Lett.}   {\bf B}}
\def\PRL{\em Phys. Rev. Lett.}
\def\PRD{{\em Phys. Rev.}    {\bf D}}
\def\ZPC{{\em Z. Phys.}      {\bf C}}
\def\EJC{{\em Eur. Phys. J.} {\bf C}}
\def\CPC{\em Comp. Phys. Commun.}

\def\white{
}
\def\black{
}

\begin{titlepage}

\begin{figure}[!t]
DESY 04--038 \hfill ISSN 0418--9833\\ 
March 2004 
\end{figure}
\bigskip

\vspace*{2cm}

\begin{center}
\begin{Large}

{\bf Evidence for a Narrow Anti-Charmed Baryon State}

\vspace*{2cm}

H1 Collaboration

\end{Large}
\end{center}

\vspace*{2cm}

\begin{abstract}
\noindent
A narrow resonance in
$D^{* \, -} p$ and $D^{* \, +} \bar{p}$ invariant mass combinations
is observed in inelastic electron-proton collisions
at centre-of-mass energies of $300 \ {\rm GeV}$ and $320 \ {\rm GeV}$ at HERA. 
The resonance has a 
mass of $3099 \pm 3 \ {\rm (stat.)} \ \pm 5 \ {\rm (syst.)} \ {\rm MeV}$
and a measured Gaussian width of 
$12 \pm 3 \ {\rm (stat.)} \ {\rm MeV}$, compatible with the
experimental resolution.
The resonance is interpreted as an anti-charmed baryon with a
minimal constituent quark composition of $uudd \bar{c}$, together with
the charge conjugate.
\end{abstract}

\vspace*{1.5cm}

\begin{center}
To be submitted to Phys. Lett. {\bf B}
\end{center}

\end{titlepage}

\begin{flushleft}

A.~Aktas$^{10}$,               
V.~Andreev$^{26}$,             
T.~Anthonis$^{4}$,             
A.~Asmone$^{33}$,              
A.~Babaev$^{25}$,              
S.~Backovic$^{37}$,            
J.~B\"ahr$^{37}$,              
P.~Baranov$^{26}$,             
E.~Barrelet$^{30}$,            
W.~Bartel$^{10}$
% ...

\bigskip{\it
% ...
$ ^{4}$ Inter-University Institute for High Energies ULB-VUB, Brussels;
          Universiteit Antwerpen (UIA), Antwerpen; Belgium$^{ c}$ \\
% ...
$ ^{10}$ DESY, Hamburg, Germany \\
% ...
$ ^{25}$ Institute for Theoretical and Experimental Physics,
          Moscow, Russia$^{ l}$ \\
 $ ^{26}$ Lebedev Physical Institute, Moscow, Russia$^{ e}$ \\
% ...
$ ^{30}$ LPNHE, Universit\'{e}s Paris VI and VII, IN2P3-CNRS,
          Paris, France \\
% ...
$ ^{33}$ Dipartimento di Fisica Universit\`a di Roma Tre
          and INFN Roma~3, Roma, Italy \\
% ...
$ ^{37}$ DESY, Zeuthen, Germany \\
% ...
$ ^c$ Supported by FNRS-FWO-Vlaanderen, IISN-IIKW and IWT \\
% ...
$ ^e$ Supported by the Deutsche Forschungsgemeinschaft \\
% ...
$ ^l$ Partially Supported by Russian Foundation
      for Basic Research, grant    no. 00-15-96584 \\
}

\vspace{2cm}

\begin{center}
  Dedicated to the memory of our dear 
friend and colleague, Ralf Gerhards.
\end{center}

\end{flushleft}

\newpage

\section{Introduction}

Several experiments have recently reported
the observation of a narrow resonance
with mass in the region of $1540 \ {\rm MeV}$,
decaying to $K^+ n$ or $K^0_s p$ \cite{strangepq}. 
This state has both baryon number and
strangeness of $+1$, such that its minimal composition
in the constituent quark model
is $uudd \bar{s}$. It has thus been interpreted
as a pentaquark \cite{qcdpq,diakonov}, the $\theta^+$. 
There is also evidence for two related states
with strangeness of $-2$ \cite{na49}.
Various models have been put forward to explain the
nature of these states and the 
structure of the multiplet that contains 
them \cite{diakonov,theory2,theory1}. 
The possibility of pentaquark states in the charm sector has also been 
discussed \cite{cspq}, with
renewed theoretical
interest in calculating their expected 
properties \cite{theory2,cheung} following
the observation of strange pentaquarks.

The electron-proton collider, HERA, is a copious producer of both charm 
and anti-charm quarks, the dominant production mechanism being 
boson-gluon fusion, $\gamma^{(\star)} g \rightarrow c \bar{c}$.
The spectroscopy of charmed hadrons can be studied using the final states
to which the quarks and anti-quarks hadronise. 
This paper reports the first evidence for a baryon 
with exotic quantum numbers in the charm sector, 
using deep inelastic scattering (DIS) data taken with the H1 detector. 
A resonance is observed when
combining 
$D^{* \, -} \rightarrow \bar{D^0} \pi^-_s 
\rightarrow K^+ \pi^- \pi^-_s$ candidates
with proton candidates and when combining 
$D^{* \, +} \rightarrow D^0 \pi^+_s \rightarrow K^- \pi^+ \pi^+_s$
candidates with antiproton 
candidates.\footnote{In the remainder of this
paper, particle charges are not generally given. Both charge 
conjugate configurations are always implied, unless explicitly
stated otherwise.
The notation $\pi_s$ is 
used to distinguish the
low momentum pion released in the $D^*$ decay from that from the $D^0$
decay.}
The resonance is also observed in an 
independent photoproduction data sample.

\section{Experimental Procedure}
\label{method}

\subsection{H1 Apparatus}
\label{detector}

The tracks from charged particles 
used in this analysis are reconstructed in the H1
central tracker, whose main components are two cylindrical drift chambers,
the inner and outer central jet chambers (CJCs), 
covering the polar angle 
region\footnote{The H1 experiment uses a coordinate system in
which the positive $z$-axis is defined by the direction of the outgoing
proton beam. The polar angle $\theta$ of a particle is defined relative
to this axis and is related to
the pseudorapidity $\eta$ by $\eta = - \ln \tan \theta / 2$.}
$20^\circ < \theta < 160^\circ$. 
The inner and outer CJCs 
are mounted 
concentrically around the beam-line,
have $24$ and $32$ sense wires,
respectively, 
and cover radii between $20 \ {\rm cm}$ and $84 \ {\rm cm}$.
The information from the CJC sense wires is digitised using $100 \ {\rm MHz}$
FADCs, providing simultaneous charge and timing measurements. 
The CJCs lie within a homogeneous magnetic field of 
$1.15 \ {\rm T}$, which allows measurements of 
the transverse momenta of charged particles.
Two additional drift chambers complement
the CJCs by precisely
measuring the $z$ coordinates of track segments and hence assist in the
determination of the particle's polar angle. 
The Central Silicon Tracker, consisting of two layers at radii of
$6 \ {\rm cm}$ and $10 \ {\rm cm}$, is also used to improve 
the charged particle track and event vertex reconstruction.
The transverse momentum resolution of the central tracker is
$\sigma(\pt) / \pt \simeq 0.005 \ \pt \ [{\rm GeV}] \ \oplus 0.015$.
The charge misidentification probability is 
negligible
for particles originating from the primary vertex which have transverse momenta
in the range relevant to this analysis. 

The specific ionisation
energy loss of charged particles is derived from the mean of the
inverse square-root
of the charge collected by all CJC sense wires with a signal above
threshold. The resolution is
$\sigma(\dedx) / (\dedx) \simeq 8 \%$ on average 
for minimum ionising particles \cite{steinhart}.

A lead/scintillating-fibre spaghetti calorimeter (SpaCal) 
is located in the direction of the outgoing 
electron beam.
It contains both electromagnetic and hadronic sections and
is used to detect the scattered electron in DIS events.
The global properties of the hadronic final state are reconstructed 
using an algorithm which takes information from the
central tracker, the SpaCal, and also from a Liquid Argon  
calorimeter, which surrounds the central tracker. 
The DIS events studied in this paper are triggered 
on the basis of 
a scattered electron in the
SpaCal, complemented by the
signals in the CJCs and 
multi-wire proportional chambers in the central tracker.
Further details of the H1 detector can be found in \cite{h1det}. 

\subsection{The DIS Data Sample}

The analysis is carried out using data taken in the years
1996-2000, when HERA collided 
electrons\footnote{The analysis uses data from periods when
the beam lepton was either a positron ($88 \%$ of the total) or an electron 
($12 \%$ of the total).}  
of energy $27.6 \ {\rm GeV}$ with protons at 
$820 \ {\rm GeV}$ (1996-1997) and $920 \ {\rm GeV}$ (1998-2000).
The integrated luminosity of the sample is $75 \ {\rm pb^{-1}}$.

The scattered
electron energy, measured in the SpaCal, 
is required to be above $8 \ {\rm GeV}$, and 
the virtuality of the exchanged 
photon\footnote{The inclusive DIS kinematic variables are defined as
$Q^2 = - q^2$, $y = q \cdot p \, / \, k \cdot p$ and
$x = -q^2 \, / \, 2 q \cdot p$, where $q$, $k$ and $p$
are the 4-vectors of the exchanged photon, the incident electron and
the incident proton, respectively.} 
is required 
to lie in the range $1 < Q^2 < 100 \ {\rm GeV^2}$, as
reconstructed from the energy and polar angle of the 
electron.
To ensure that the
hadronic final state lies in the central region of the detector,
the inelasticity of the event 
is required to satisfy $0.05 < y < 0.7$,
calculated using the scattered electron kinematics. 
The $z$ coordinate of the
event vertex, reconstructed using the central tracker,
is required to lie within $35 \ {\rm cm}$ 
of the mean position for $ep$ interactions. 
The difference between the total
energy $E$ and the longitudinal component of the total
momentum $p_z$, calculated 
from the electron and the hadronic final state, is restricted
to $E - p_z > 35 \ {\rm GeV}$. This requirement suppresses  
photoproduction background, where a hadron
fakes the electron signature.

\subsection{Selection of {\boldmath $D^*$} Meson and Proton Candidates}
\label{dsprec}

The decay channel $D^* \rightarrow D^0 \pi_s \rightarrow K \pi \pi_s$
is used to reconstruct $D^*$ mesons.
The charged particle selection criteria, $\dedx$ requirements
and transverse momentum cuts on the decay products are 
very similar to those used in previous H1 analyses \cite{f2c}.
Unlike-charge particle combinations are made to form 
$K^{\mp} \pi^{\pm}$ pairs, where the 
particles are required to be consistent 
with kaons and pions according to their $\dedx$ measurements.
Those combinations that give rise to an
invariant mass within $60 \ {\rm MeV}$ of the nominal $D^0$ mass
of $1864.5 \ {\rm MeV}$ \cite{pdg} 
are combined with further pion candidates
($\pi_s$) with opposite charge to the kaon.
To obtain good experimental resolution 
and background rejection in the $D^*$
reconstruction, the standard mass difference technique \cite{deltam} is
used, based on the variable
\begin{eqnarray} 
\Delta M_{D^*} = m (K \pi \pi_s) -  m (K \pi) \ .
\label{mddef}
\end{eqnarray}
The sample is 
restricted to a region where the background to the
$D^*$ signal is relatively
small by requiring that the $D^*$ candidates
have transverse momentum $\pt(D^*) > 1.5 \ {\rm GeV}$,
pseudorapidity $-1.5 < \eta(D^*) < 1$ and production
elasticity
$z(D^*) = (E - p_z)_{D^*} / 2 y E_e > 0.2$, where $E_e$ is the electron beam
energy.

\begin{figure}[p] \unitlength 1mm
 \begin{center}
 \begin{picture}(100,180)
  \put(-10,90){\epsfig{file=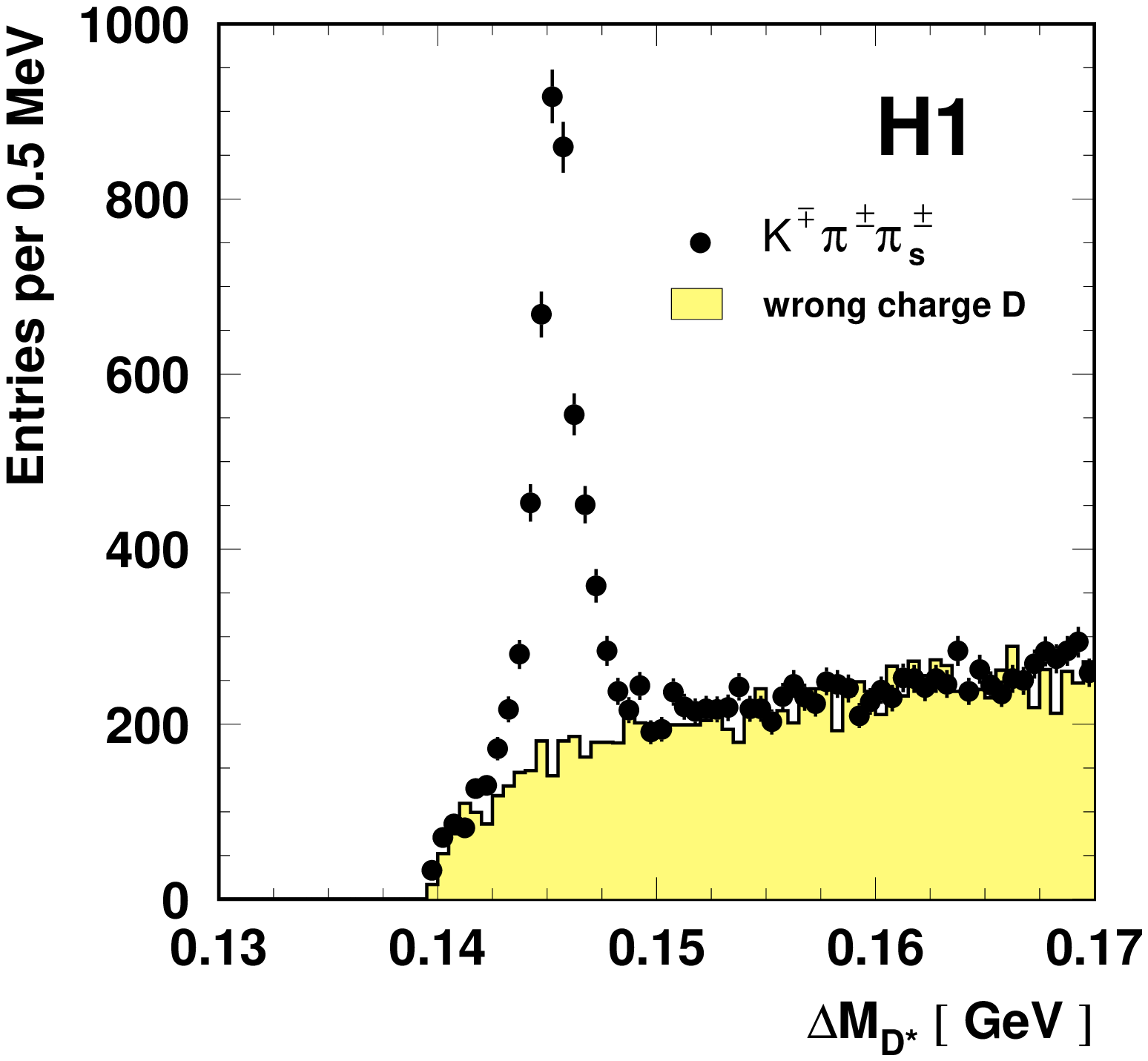,width=0.8\textwidth}}
  \put(-10,-10){\epsfig{file=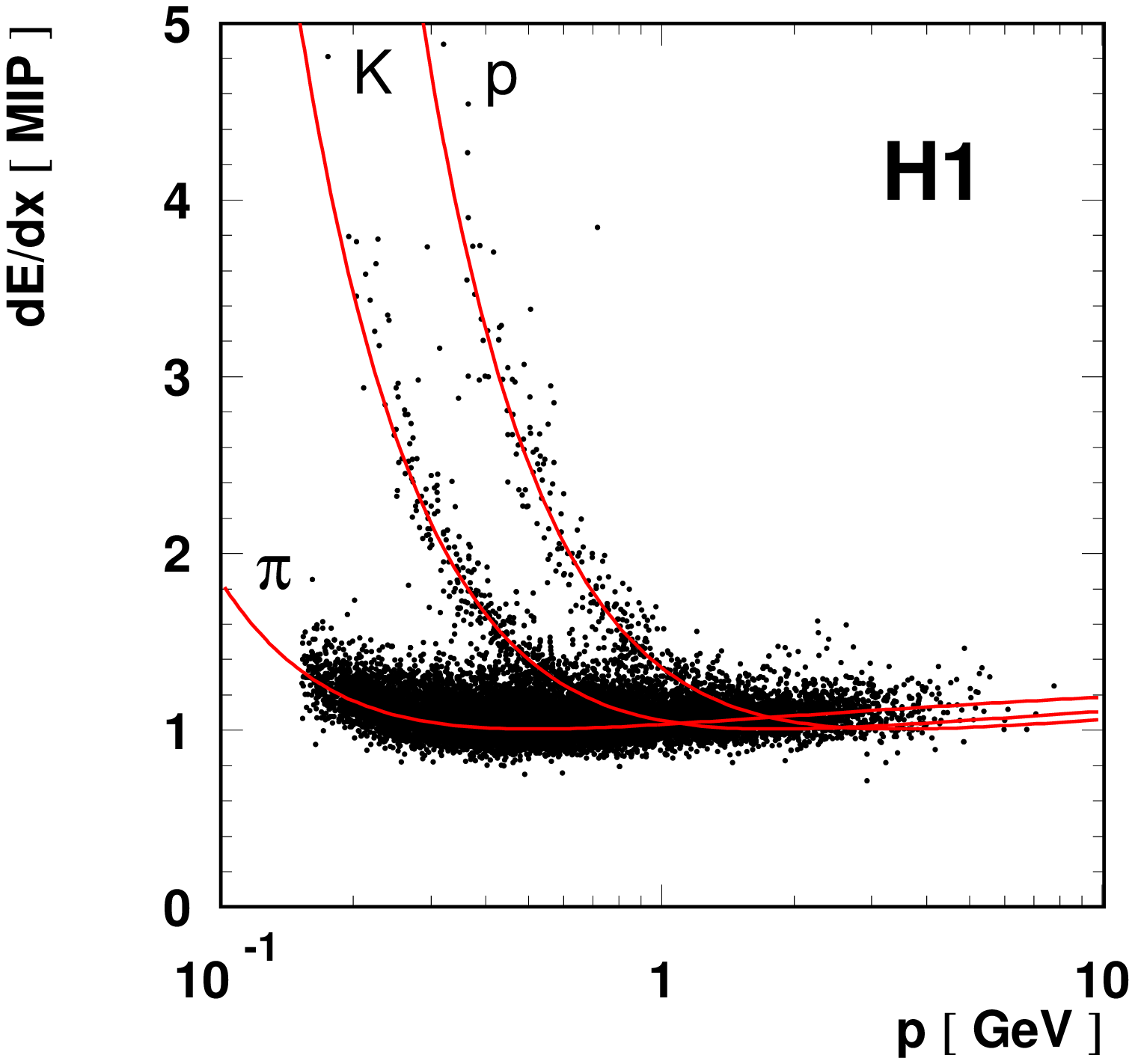,width=0.8\textwidth}}
  \put(18,122){{\Large {\bf \fontfamily{phv}\selectfont (a)}}}  
  \put(18,22){{\Large {\bf \fontfamily{phv}\selectfont (b)}}}
 \end{picture}
 \end{center}
\caption{(a) $\Delta M_{D^*}$ distribution for
$K^\mp \pi^\pm \pi_s^\pm$ combinations  
as described in the text. For comparison, the distribution from
``wrong charge $D$'' combinations, where the $K$ and $\pi$ yielding the $D^0$
mass have the same charge, is also shown.
(b) Specific ionisation energy loss relative to that of a minimally ionising
particle, plotted against momentum, for the sample described
in the text.
The curves indicate parameterisations of the most probable responses of the 
CJCs for pions, kaons and protons.}
\label{dstardis} 
\end{figure}

The resulting $\Delta M_{D^*}$ distribution 
is shown in figure~\ref{dstardis}a.
Here and elsewhere in this paper, the error bars shown represent 
the square-root of the numbers of entries in each bin.  
A prominent
signal on a smooth background is observed around the expected $D^* - D^0$ 
mass difference. 
The $\Delta M_{D^*}$ 
distribution is also shown in figure~\ref{dstardis}a for 
a ``wrong charge $D$'' background sample, which is formed by $K^\pm \pi^\pm$
combinations in the accepted $D^0$ mass range. 
The ``wrong charge $D$'' distribution gives a good description of the 
correct-charge $D^0$ combinations
away from the $D^*$ peak. 
Candidate $D^*$ mesons for which $\Delta M_{D^*}$ lies within
$\pm 2.5 \ {\rm MeV}$ of the nominal mass difference 
$m(D^*) - m(D^0) = 145.4 \ {\rm MeV}$ \cite{pdg} 
are selected for further analysis. 

Proton candidates are selected using requirements on the 
particle $\dedx$ measurements.
Figure~\ref{dstardis}b shows the $\dedx$ values, 
plotted against momentum,
for a sample of particles which yield a mass
$M(D^* p) < 3.9 \ {\rm GeV}$ 
(see equation~\ref{mdpdef}) when combined with 
$K \pi \pi_s$ candidates falling in the accepted $\Delta M_{D^*}$ region.
The likelihoods that a particle is a pion, kaon or proton
are obtained from the proximity of the 
measured $\dedx$ to the most probable values for each particle type
at the reconstructed momentum. The most probable $\dedx$ values 
are derived from 
phenomenological parameterisations \cite{steinhart}, shown in 
figure~\ref{dstardis}b, which are
based on the Bethe-Bloch formula. 
The normalised
proton likelihood $L_p$ is defined to be the ratio of the proton likelihood
to the sum of the pion, kaon and proton likelihoods.
For momenta $p(p) < 2 \ {\rm GeV}$, a requirement $L_p > 0.3$
is applied, which selects protons where they are clearly identified at
low momentum and suppresses contributions close to the crossing points
of the proton, pion and kaon parameterisations.
For $p(p) > 2 \ {\rm GeV}$, 
the requirement is loosened to $L_p > 0.1$, which suppresses background from
particles with large $\dedx$ such as electrons.
The main selection criteria are summarised in table~\ref{cuts}.

\begin{table}[h]
  \begin{center}
    \begin{tabular}{|l|l|}
      \hline

{\boldmath $D^0$}     & $\pt(K) > 500 \ {\rm MeV}$               \\
                      & $\pt(\pi) > 250 \ {\rm MeV}$             \\
                      & $\pt(K) + \pt(\pi) > 2 \ {\rm GeV}$      \\
                      & $|m(K \pi) - m(D^0)| < 60 \ {\rm MeV}$   \\ \hline
{\boldmath $D^*$}     & $\pt(\pi_s) > 120 \ {\rm MeV}$           \\
                      & $|\Delta M_{D^*} - m(D^*) + m(D^0)| < 2.5 \ {\rm MeV}$ 
                                                     \\
                      & $\pt(D^*) > 1.5 \ {\rm GeV}$             \\ 
                      & $-1.5 < \eta(D^*) < 1$                   \\
                      & $z(D^*) > 0.2$                           \\ \hline
{\boldmath $p$}       & $\pt(p) > 120 \ {\rm MeV}$               \\
                      & $L_p > 0.3$ for $p(p) < 2 \ {\rm GeV}$   \\ 
                      & $L_p > 0.1$ for $p(p) > 2 \ {\rm GeV}$   \\ \hline
    \end{tabular}
    \caption{Summary of the kinematic and proton energy loss
selection criteria
applied to define the $D^*$ and proton candidates.}
  \label{cuts}
  \end{center}
\end{table}

\section{Analysis of {\boldmath $D^* p$} Combinations}
\label{dstarp}

\subsection{{\boldmath $D^* p$} Invariant Mass Distributions}
\label{massdist}

\begin{figure}[p] \unitlength 1mm
 \begin{center}
 \begin{picture}(100,180)
  \put(-25,88){\epsfig{file=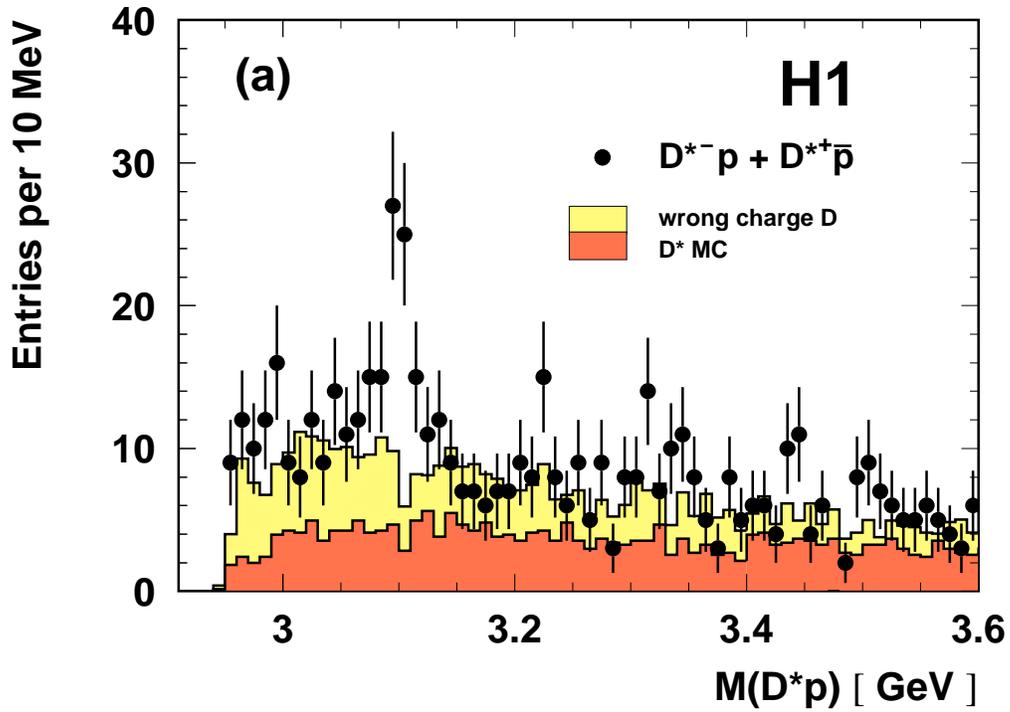,width=0.95\textwidth}}
  \put(-25,-15){\epsfig{file=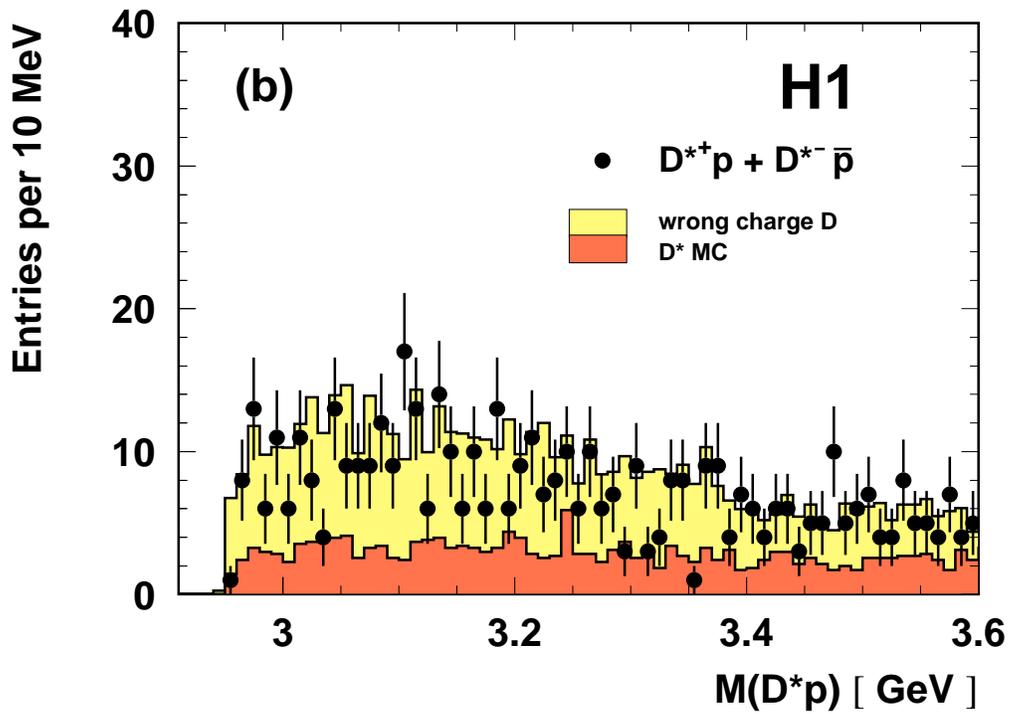,width=0.95\textwidth}}
  \put(13,185){{\Large {\bf \fontfamily{phv}\selectfont (a)}}}  
  \put(13,81){{\Large {\bf \fontfamily{phv}\selectfont (b)}}}
 \end{picture}
 \end{center}
\caption{Distributions in $M(D^* p)$ for
(a) opposite-charge and (b) same-charge $D^* p$ combinations. The data are
compared with a two-component background model in which 
``wrong charge $D$'' $K^\pm \pi^\pm$ combinations 
are used to describe non-charm related
background and the ``$D^*$ MC'' simulation describes background involving
real $D^*$ mesons.}
\label{signal}
\end{figure}

The $D^*$ and proton candidates are combined to form the
mass difference 
$m(K \pi \pi_s p) - m(K \pi \pi_s)$, to which the
$D^*$ mass of $2010.0 \ {\rm MeV}$ \cite{pdg} 
is added to obtain the
mass of the $D^* p$ combination. The distributions in
\begin{eqnarray} 
M(D^* p) = m(K \pi \pi_s p) - m(K \pi \pi_s) + m(D^*)
\label{mdpdef}
\end{eqnarray} 
are shown in figure~\ref{signal}a 
for ``opposite-charge'' $D^* p$
combinations ($K^- \pi^+ \pi_s^+ \bar{p}$ and 
$K^+ \pi^- \pi_s^- p$) and in figure~\ref{signal}b  
for ``same-charge'' $D^* p$ combinations 
($K^- \pi^+ \pi_s^+ p$ and 
$K^+ \pi^- \pi_s^- \bar{p}$). 
A clear and
narrow peak is observed for the opposite-charge combinations at
$M(D^* p) \simeq 3100 \ {\rm MeV}$.
Approximately half of the events in this signal arise from each of the 
$D^{* \, -} p$ and $D^{* \, +} \bar{p}$ combinations
(see section~\ref{signif}). The distribution for 
the same-charge combinations
shows 
a small enhancement in the 
$M(D^* p)$ region in which the opposite-charge 
signal is observed.

The background distributions for the $D^* p$ combinations
are modelled by the sum of two contributions, which are shown
in figure~\ref{signal}.
Background from random combinations not involving charm is
modelled using the ``wrong charge $D$'' combinations,  
as described in section~\ref{dsprec}, combined with proton candidates
as for the correct-charge $D^0$ sample. Combinatorial background from
$D^*$ mesons with
real or misidentified protons is modelled using simulated events from 
the RAPGAP \cite{rapgap} Monte Carlo model applied to 
$D^*$ production in DIS, 
including string fragmentation and decays from 
JETSET \cite{lund,jetset}. The RAPGAP model 
gives a good description of the shapes of the inclusive $D^*$ distributions.
This contribution (``$D^*$ MC'' in figure~\ref{signal})
is normalised according to the $D^*$ yield in the 
data (figure~\ref{dstardis}a). 

No significant structures are observed in either component
of this background model. 
The model gives a 
reasonable description of the shape and normalisation of the 
$M(D^* p)$ distribution away from the signal 
region for the opposite-charge combinations. The $M(D^* p)$ distribution
from the same-charge combinations is also well described in shape, though
the model prediction 
lies approximately $15 \%$ above the data.

Alternative models have been studied for the background distribution 
for the opposite-charge $D^* p$ combinations. 
Similar distributions to those shown in figure~\ref{signal} are obtained
when a DJANGO \cite{django} simulation of inclusive DIS is used to
replace both model components. 
The same is true when the RAPGAP 
model of the $D^*$-related background is replaced by
simulations with modified 
parton shower dynamics (CASCADE \cite{cascade}) or fragmentation
(HERWIG \cite{herwig}). In all cases, no resonant structures are observed in 
the simulated $M(D^* p)$ distributions. 
Possible contributions from beauty decays have been considered
using a further RAPGAP Monte Carlo simulation. 
After normalising to
the luminosity of the data, the resulting contribution is negligible. 

The events giving contributions in the peak region
of the opposite-charge $M(D^* p)$ distribution 
have been visually scanned and no  
anomalies are observed in the events or the candidate tracks.
All entries within $\pm 24 \ {\rm MeV}$ of
the peak arise from different events, with one
exception where the same $\pi$, $\pi_s$ and proton candidates 
contribute with two different $K$ candidates.
For the full $M(D^* p)$ range shown in figure~\ref{signal}a, there are
an average of $1.12$ entries per event.
The signal is present in each case when 
the data are divided into two sub-samples of similar size, discriminated in
variables such as $x$ or $Q^2$, the
pseudorapidity or transverse momentum
of the $D^* p$ composite, or the data taking period.  
The peak also remains clearly visible for 
reasonable variations in the binning or selection 
criteria. 
In all cases, the observed mass and width of the peak
are stable to within a few ${\rm MeV}$.

\subsection{Particle Identification Tests}
\label{particleid}

\begin{figure}[h]
\center
 \begin{picture}(100,120)
  \put(-22,-15){\epsfig{file=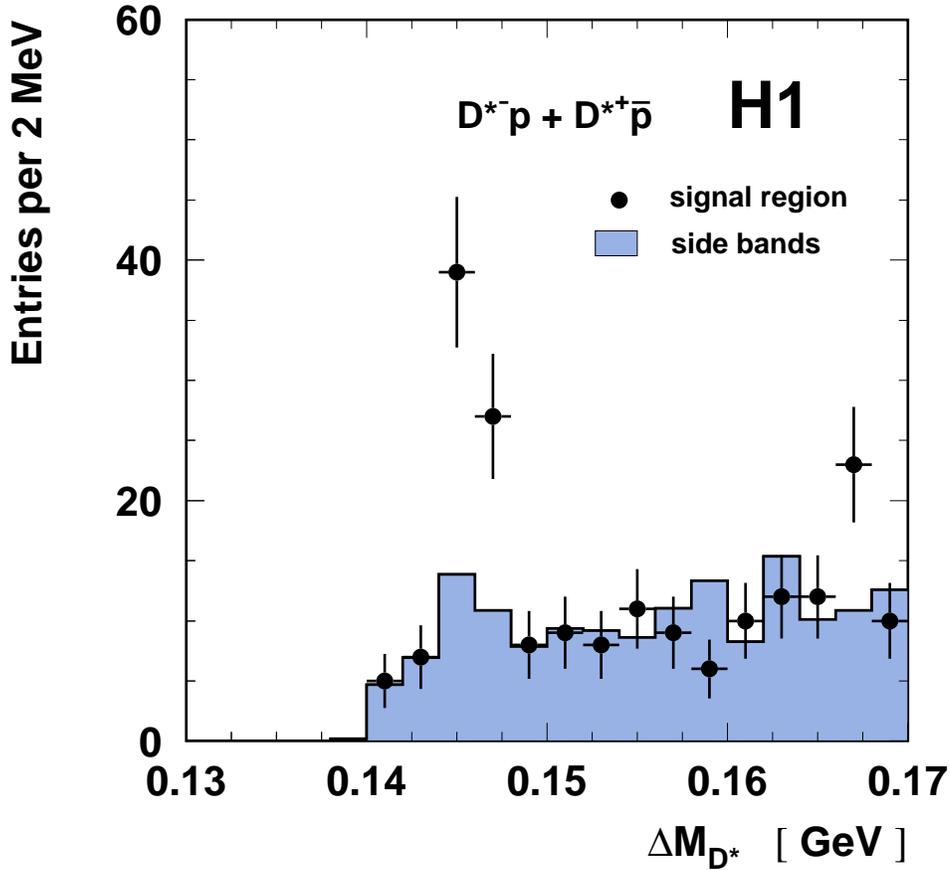,width=\textwidth}}
 \end{picture}
\caption{$\Delta M_{D^*}$ distribution for events 
in a $30 \ {\rm MeV}$ window about the signal in the
opposite-charge $M(D^* p)$ distribution,
with no requirement on $\Delta M_{D^*}$, 
compared with
the corresponding distribution from side bands in the  
$M(D^* p)$ distribution, normalised according to the 
widths of the chosen sample regions.}
\label{backwards} 
\end{figure}

The $D^*$ and proton content of the
signal in the mass distribution from
opposite-charge $D^* p$ combinations has been investigated in
complementary studies. The $D^*$ content is tested
by forming the $M(D^* p)$ distribution (equation~\ref{mdpdef}) with the full
proton selection, but with no requirement on $\Delta M_{D^*}$. The
$\Delta M_{D^*}$ distribution (equation~\ref{mddef})
is shown in figure~\ref{backwards}
for events in a $\pm 15 \ {\rm MeV}$ window around the $D^* p$ signal
($3085 < M(D^* p) < 3115 \ {\rm MeV}$). 
For comparison, a similar distribution
is shown, taken from side bands with 
$2990 < M(D^* p) < 3070 \ {\rm MeV}$ and 
$3130 < M(D^* p) < 3210 \ {\rm MeV}$, scaled by a factor
of $3/16$ to account for the different widths of the sample regions. 
Away from the $D^*$ peak,
the distribution in $\Delta M_{D^*}$
from the $M(D^* p)$ side bands gives a good description
of that from the $M(D^* p)$ signal region, in both
shape and normalisation. However, there is a clear difference
around the expected value of $\Delta M_{D^*}$ for 
true $D^*$ mesons,
where the distribution from the signal region lies well above that
from the side bands. The signal region in $M(D^* p)$
is thus significantly richer in $D^*$ mesons than is the case elsewhere in
the distribution. 

\begin{figure}[h]
\begin{center}
 \begin{picture}(100,90)
  \put(-25,-15){\epsfig{file=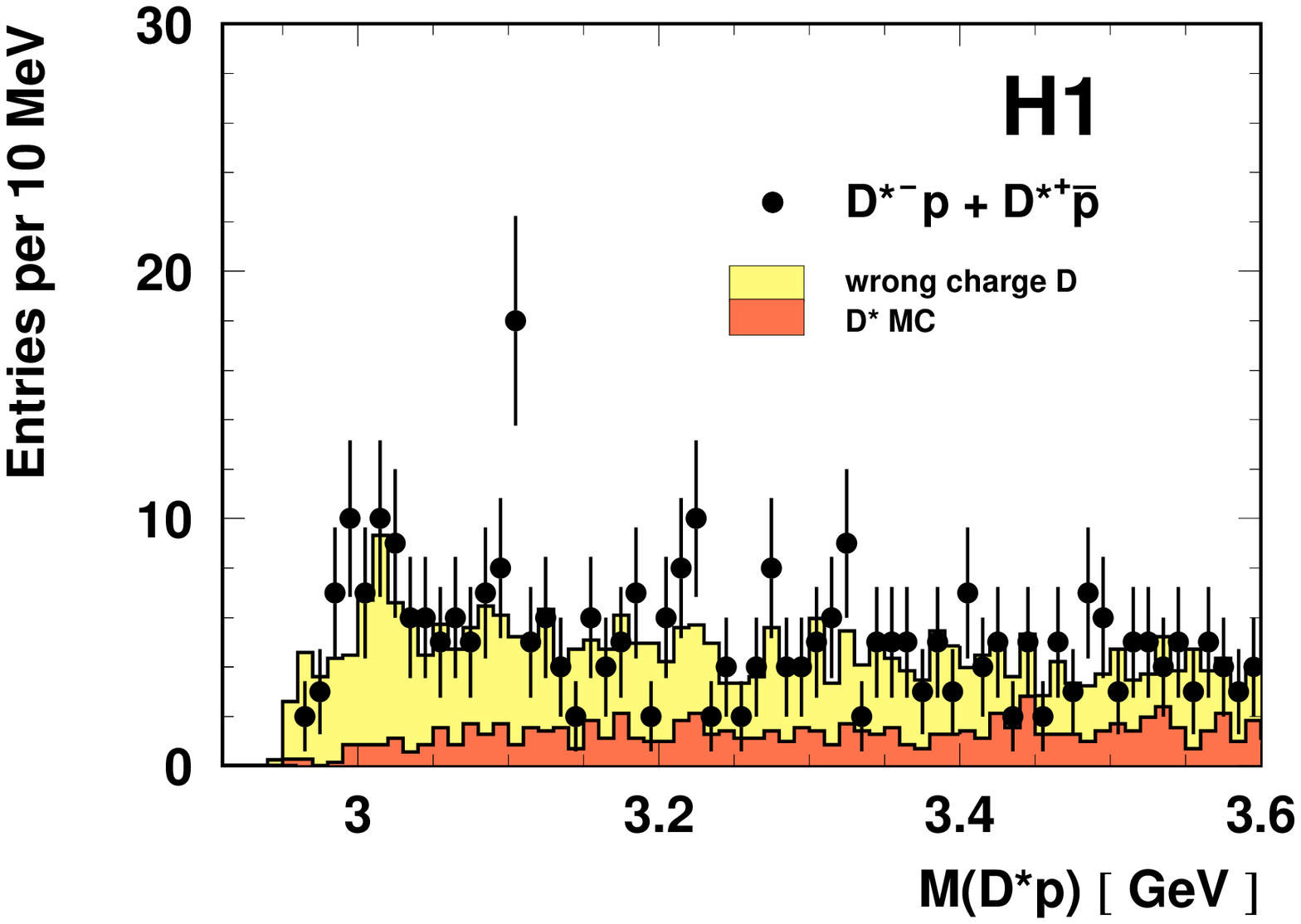,width=0.95\textwidth}}
 \end{picture}
\end{center}
\caption{$M(D^* p)$ distribution for a low momentum proton
selection with $p(p) < 1.2 \ {\rm GeV}$. The data are
compared with a two-component background model in which 
``wrong charge $D$'' $K^\pm \pi^\pm$ combinations 
are used to describe non-charm related
background and the ``$D^*$ MC'' simulation describes background involving
real $D^*$ mesons.}
\label{lowp} 
\end{figure}

The protons are clearly identified at low momentum, where the most
probable $\dedx$ 
for protons is well separated from those for other 
particle species. 
The analysis has been repeated with the
proton momentum restricted to $p(p) < 1.2 \ {\rm GeV}$, the measured
$\dedx$ required to be larger than that for
a minimum ionising particle by a factor of at least $1.15$ and the
proton likelihood requirement modified to $L_p > 0.5$.
With this tighter proton selection,  
the requirements on $z(D^*)$ and 
$\pt(K) + \pt(\pi)$ are removed. In figure~\ref{lowp},
the $M(D^* p)$
distribution for this selection
is compared with the predictions of the background model 
described in section~\ref{massdist}. The 
enhancement in the region $M(D^* p) \simeq 3100 \ {\rm MeV}$ 
remains visible. The candidate proton tracks in the signal region
($3085 < M(D^* p) < 3115 \ {\rm MeV}$) have an average 
$\langle L_p \rangle = 0.92$.

Several further checks of the particle identification
have been carried out using the data. 
No signal is obtained when the
$D^*$ selection is modified such that the ``wrong charge $D$'' candidates
are taken (figure~\ref{signal}a) or when
the $D^*$ candidates are replaced by 
$K \pi \pi_s$
combinations from a side band of the $\Delta M_{D^*}$ distribution.
No clear peaks are observed when the proton candidates are assumed
to have the kaon or pion mass. 
The signal is also not observed when the proton mass hypothesis 
is retained, but the $\dedx$ requirement 
is modified to select pion or kaon candidates
rather than proton candidates. 

\subsection{Kinematic and Reflection Tests}
\label{kinematic}

Possible kinematic reflections that could fake the signal have been ruled
out by studying invariant mass distributions 
and correlations involving the $K$, $\pi$, 
$\pi_s$ and proton candidates under various particle mass hypotheses. 
For example, 
there is no evidence for any resonant structure or correlations with
$M(D^* p)$
in the invariant mass combinations 
$m(Kp)$, $m(\pi p)$ or $m(\pi_s p)$ of the 
proton candidates with the decay products
of the $D^*$ meson.

Detailed studies have been carried out of the contribution to the 
$M(D^* p)$ distribution from the neutral, orbitally 
excited, P-wave $D^0_1 (2420)$ and $D^{0 \, *}_2 (2460)$ mesons 
and their charge conjugates \cite{pdg}, 
both of which decay to $D^{* \, \pm} \pi^\mp$.
A simulation of the $D^0_1 \rightarrow D^* \pi$ and 
$D^{0 \, *}_2 \rightarrow D^* \pi$ decays 
with the PYTHIA \cite{pythia} Monte Carlo generator
is used to estimate their
contribution to the observed signal.
The simulated widths are
set to the results from recent measurements \cite{belle}
and the normalisation 
is obtained from the observed $D^0_1$ and $D^{0 \, *}_2$ yields 
in the data, as
obtained from the distribution in 
$M(D^* \pi) = m(K^\mp \pi^\pm \pi_s^\pm \pi^\mp) - 
m(K^\mp \pi^\pm \pi_s^\pm) + m(D^*)$.
The reflections due to the $D^0_1$ and $D^{0 \, *}_2$ mesons 
when the decay pion is misidentified as a 
proton yield a broad distribution in $M(D^* p)$, with a maximum
below the signal region. 
The predicted contribution within $\pm 24 \ {\rm MeV}$ of the observed peak
is approximately four events. 
The signal in the measured $M(D^* p)$ distribution covers the full available
phase space in $M(D^* \pi)$, with no evidence for enhancements
in the region of the
$D^0_1$ and $D^{0 \, *}_2$ mesons. 
The possibility of reflections involving the $D_{s1} (2536)$ or
$D_{sJ} (2573)$, are similarly ruled out.\footnote{Given the 
proximity of the mass of the
observed resonance 
to the $J/\psi$ mass, possible backgrounds involving $J/\psi$
decays have been considered. Baryon number and other 
conservation laws would be violated 
by the decay of the $J/\psi$ to $D^* p$. If the signal were due to $J/\psi$
decays with misidentified particles, the reconstructed mass would no longer
lie at the $J/\psi$ value.} 

\begin{figure}[p] \unitlength 1mm
 \begin{center}
 \begin{picture}(100,180)
  \put(-18,72){\epsfig{file=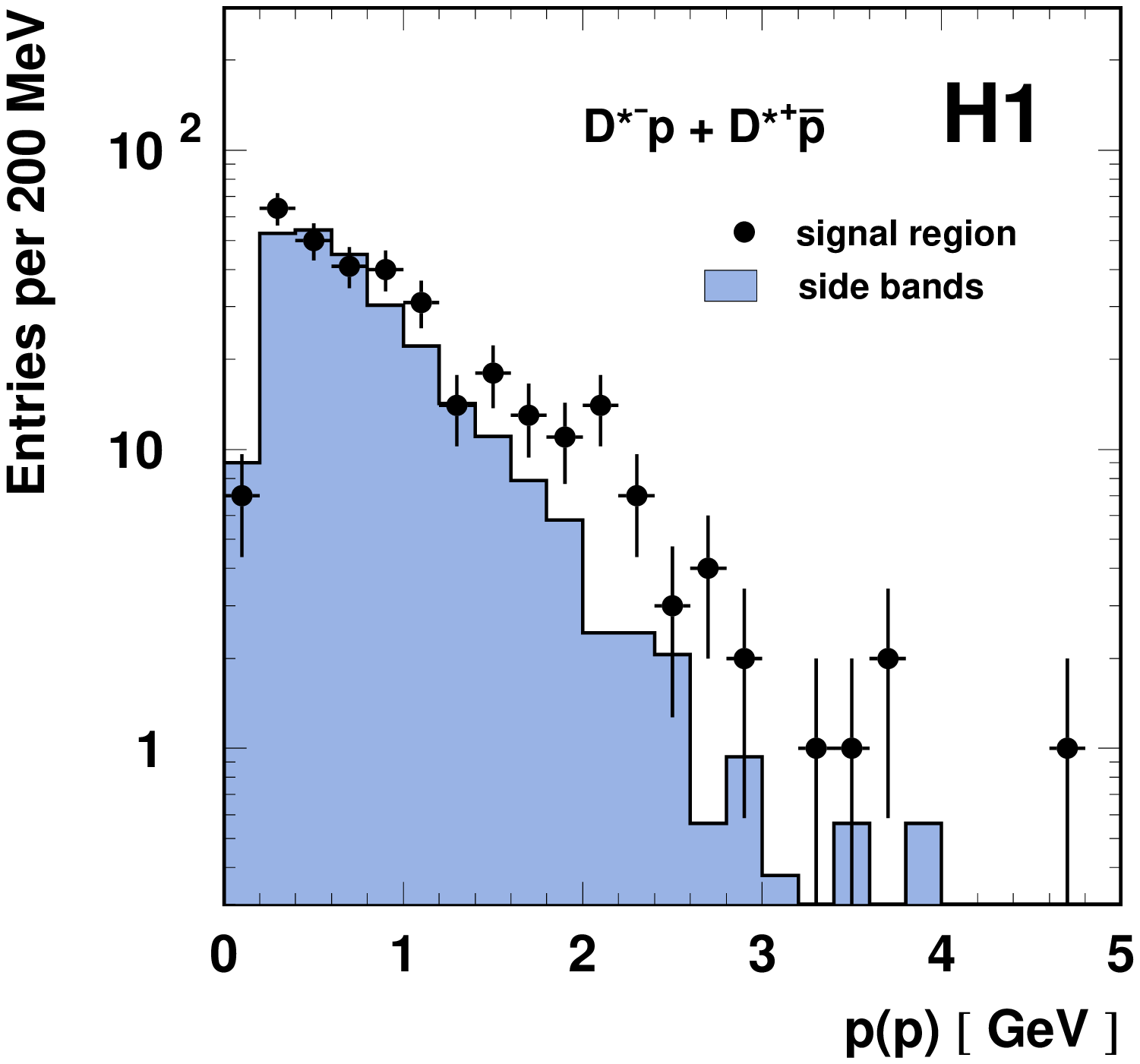,width=0.95\textwidth}}
  \put(-14.5,-15){\epsfig{file=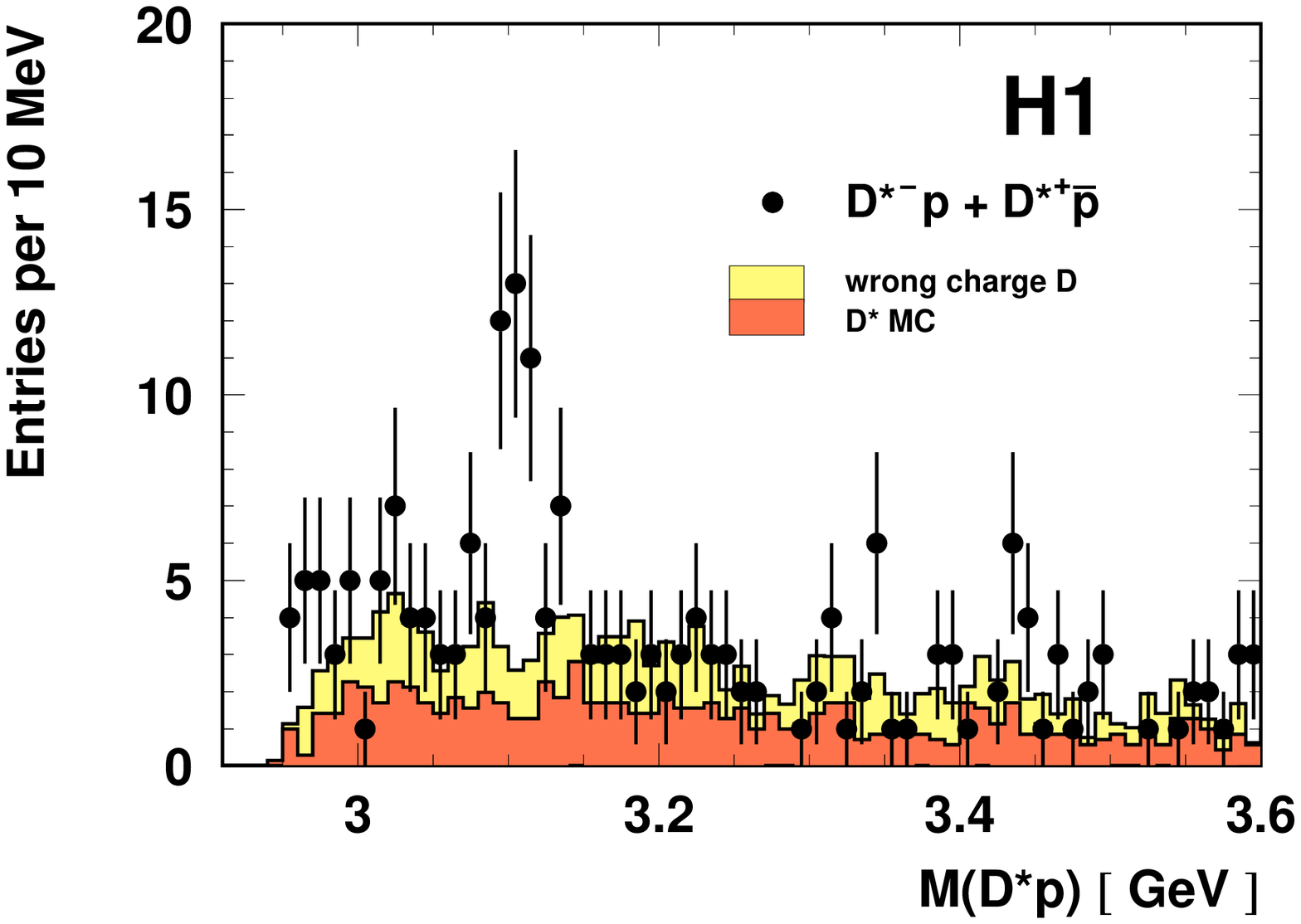,width=0.82\textwidth}}
  \put(18,185){{\Large {\bf \fontfamily{phv}\selectfont (a)}}}  
  \put(18,68){{\Large {\bf \fontfamily{phv}\selectfont (b)}}}
 \end{picture}
 \end{center}
\caption{(a) Momentum distributions for all charged particles yielding 
$M(D^* p)$ values falling in the signal and 
side band regions of $M(D^* p)$, as defined in 
section~\ref{particleid}, when combined with $D^*$ candidates of
opposite charge.
(b) $M(D^* p)$ distribution for $p(p) > 2 \ {\rm GeV}$,
with no proton $\dedx$ requirements. The data are
compared with a two-component background model in which 
``wrong charge $D$'' $K^\pm \pi^\pm$ combinations 
are used to describe non-charm related
background and the ``$D^*$ MC'' simulation describes background involving
real $D^*$ mesons.} 
\label{highp} 
\end{figure}

The kinematics of the $D^*$ and proton candidates 
from the decay of a resonance would be expected to be different 
from those 
of the background distribution. Such a difference is observed
for the opposite-charge $D^* p$ 
signal, as illustrated in figure~\ref{highp}a. The
momentum distribution $p(p)$ is shown for all particles of opposite
charge to the $D^*$ candidate which lead to entries in the signal 
and side band regions of $M(D^* p)$, as defined in
section~\ref{particleid}. No requirements are placed on the 
proton likelihood. 
The two side bands with larger and smaller $M(D^* p)$ than the signal
give rise to compatible momentum spectra, both of which are 
significantly softer
than that in the signal region.
A similar difference is
observed when the $M(D^* p)$ side band is replaced by combinations 
which lie
in the signal region of $M(D^* p)$, but fall in
a side band of the $\Delta M_{D^*}$ distribution.

Figure~\ref{highp}a suggests that, with no proton $\dedx$ requirements,
the signal-to-background ratio improves as $p(p)$
increases.
In figure~\ref{highp}b, the $M(D^* p)$ distribution
is shown for $p(p) > 2 \ {\rm GeV}$, with
no requirement on the proton likelihood. A strong
signal is observed,  
with a reduced background
which remains well described by the background
model. The peak value and width of the observed signal are 
compatible with those
for the standard selection shown in figure~\ref{signal}a.

\subsection{Photoproduction Analysis}

The analysis has also been carried out using an independent
sample provided by H1
data from the photoproduction region, where the 
scattered electron passes at a small angle into the 
backward beampipe,
implying $Q^2 \, \lapprox \, 1 \ {\rm GeV^2}$. 
The hadronic final state is used to reconstruct $y$ \cite{jb} and the
selection $0.2 < y < 0.8$ is imposed.
The combinatorial background to the $D^*$ selection is 
significantly larger 
for photoproduction than for DIS.
To compensate for this, tighter proton and $D^*$ selections are imposed.
The cut on the $D^*$ transverse momentum is modified to
$\pt(D^*) > 2 \ {\rm GeV}$.
The region $1.6 < p(p) < 2.0 \ {\rm GeV}$,
around the point at which 
the $\dedx$ parameterisations for protons and pions cross,
is excluded and the requirement $L_p > 0.25$ is made elsewhere. 

\begin{figure}[h]
\begin{center}
 \begin{picture}(100,92)
  \put(-25,-18){\epsfig{file=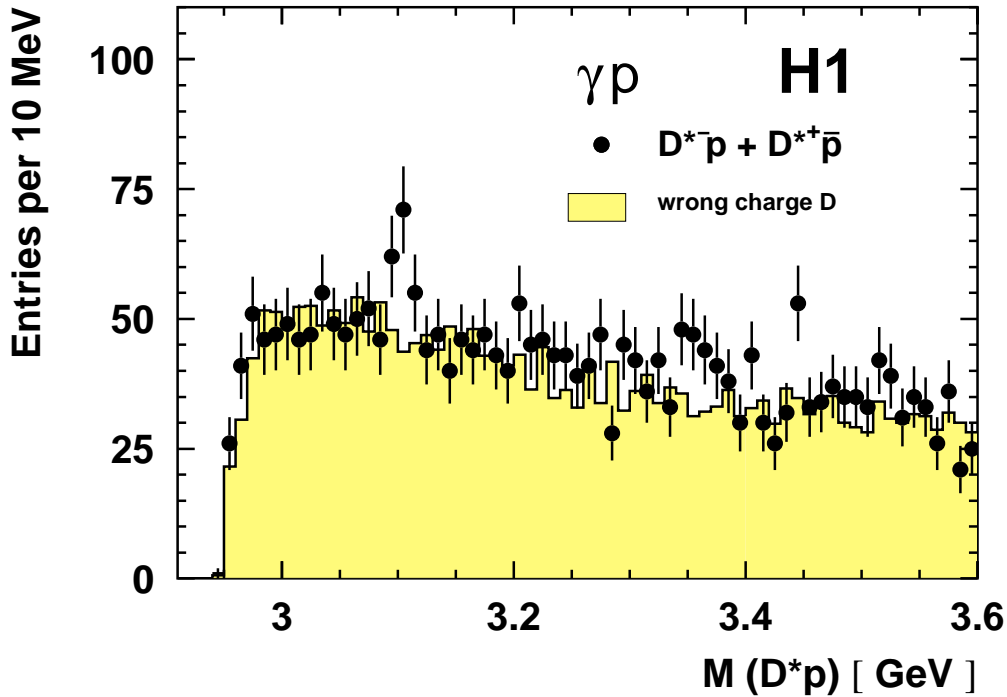,width=0.95\textwidth}}
 \end{picture}
 \end{center}
\caption{$M(D^* p)$ distribution from the
photoproduction analysis, compared with a 
background model derived from ``wrong charge $D$'' $K^\pm \pi^\pm$ 
combinations.}
\label{crosschecks} 
\end{figure}

The distribution in $M(D^* p)$ for 
opposite-charge $D^* p$ combinations in photoproduction is shown in 
figure~\ref{crosschecks}. 
Again, a clear signal is observed 
near $M(D^* p) = 3100 \ {\rm MeV}$, with mass and width compatible with
those in the DIS case.
The background distribution is reasonably 
modelled by the ``wrong charge $D$'' selection.
The photoproduction signal is 
also separately observed for $p(p) < 1.6 \ {\rm GeV}$
and for $p(p) > 2.0 \ {\rm GeV}$.

\section{Signal Assessment}
\label{signif}

Fits to the $M(D^* p)$ distribution from the
opposite-charge $D^* p$ combinations in DIS have been carried out 
to evaluate the peak position, width and statistical significance. 
Assuming the measured width to be dominated by the
experimental resolution, a Gaussian distribution
is used for the signal, with
the peak position, the
width and the normalisation 
as free parameters. The background is parameterised with
a power law of the form
$\alpha \left[ M(D^* p) - m(D^*) \right]^\beta$, with $\alpha$
and $\beta$ as free parameters. A log-likelihood 
fit is made in
the range $2950 < M(D^* p) < 3600 \ {\rm MeV}$. 

\begin{figure}[h]
\begin{center}
 \begin{picture}(100,115)
  \put(-40,-20){\epsfig{file=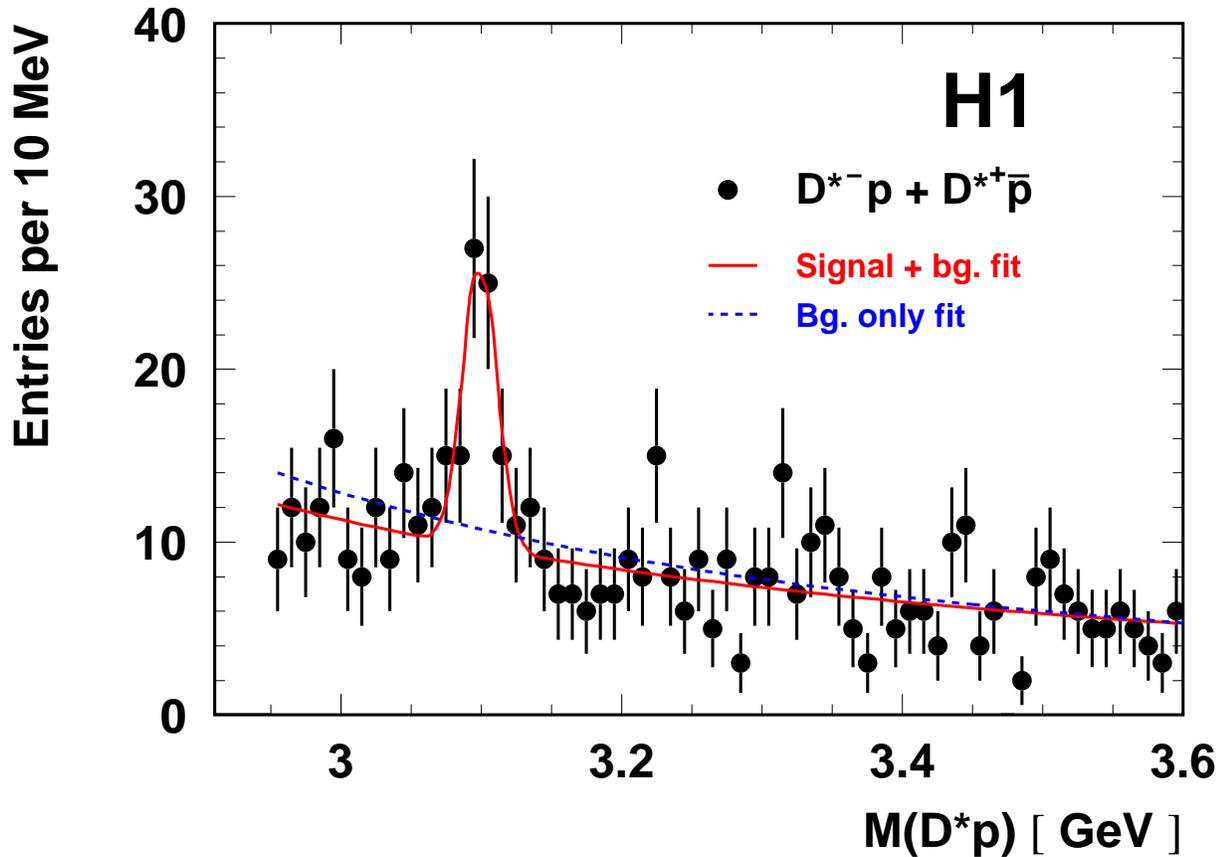,width=1.15\textwidth}}
 \end{picture}
\end{center}
\caption{$M(D^* p)$ distribution from opposite-charge
$D^* p$ combinations in DIS, compared with the results of a fit
in which both signal and background components are included (solid
line) and with the results of a fit in which only the background 
component is included (dashed line).} 
\label{final} 
\end{figure}

The results of this fit are compared
with the data in figure~\ref{final}. 
They are also summarised in table~\ref{fittab},
together with the results of separate fits to the $D^{* \, -} p$ and
$D^{* \, +} \bar{p}$ contributions.
The fit yields a
peak position of $M(D^* p) = 3099 \pm 3 \ {\rm (stat.)} 
\ {\rm MeV}$.
The root-mean-square (RMS) width of the Gaussian is 
$12 \pm 3 \ {\rm (stat.)} \ {\rm MeV}$,
compatible with the experimental resolution
of $7 \pm 2 \ {\rm MeV}$,
as determined from a simple simulation of the observed 
resonance with zero width and an isotropic decay distribution.
The signal consists of $N_s = 50.6 \pm 11.2$ events,
from which the observed $D^* p$ resonance is estimated to contribute 
roughly $1 \%$ of the total $D^*$ production rate in the kinematic
region studied. 
The fit results
are not significantly affected when the
background parameterisation is replaced with a polynomial or when the
full distribution is fitted
with the inclusion of a function to describe the rise at threshold.
The results are stable against shifts in the binning, 
changes to the bin width
in the range $1 \ {\rm MeV}$ to $20 \ {\rm MeV}$ and 
variations in the selection criteria which do not significantly alter the
signal-to-background ratios for the $D^*$ or proton candidates.
 
\begin{table}[h]
  \begin{center}
    \begin{tabular}{|l|c|c|c|}
      \hline
{\bf Sample} & {\bf Mass} & {\bf Gaussian width} & {\boldmath $N_s$} \\
             & {\bf ({\boldmath $\rm MeV$})} & {\bf ({\boldmath $\rm MeV$})} & \\ \hline  
{\boldmath $D^{* \, +} \bar{p} + D^{* \, -} p$} & $3099 \pm 3$ & $12 \pm 3$ & $50.6 \pm 11.2$  \\
{\boldmath $D^{* \, -} p$}       & $3102 \pm 3$ & $9 \pm 3$ & $25.8 \pm 7.1$  \\ 
{\boldmath $D^{* \, +} \bar{p}$} & $3096 \pm 6$ & $13 \pm 6$ & $23.4 \pm 8.6$ \\ \hline
    \end{tabular}
    \caption{Results of the fit as described in the text for 
opposite-charge $D^* p$ combinations.
The fitted position and Gaussian RMS width of the peak 
are given, together with the total number of signal events ($N_s$). 
The statistical uncertainties quoted take
account of the correlations between the variable parameters in the fit.}
  \label{fittab}
  \end{center}
\end{table}

The systematic uncertainty on the mass of the peak is
$5 \ {\rm MeV}$, estimated from the reconstructed masses of known
states, such as
the $J/\psi$,  
with decays to particles in a similar momentum range,
and from the variations in the peak position
when the fitting procedure or selection criteria are modified.

The probability that the background distribution fluctuates to 
produce the signal has
been evaluated
by comparing the observed number of events with background estimates 
for a window spanning
$3075 < M(D^* p) < 3123 \ {\rm MeV}$, corresponding to $\pm 2 \ \! \sigma$ 
about the peak position according to the fit. 
The total number of 
events in this interval is 95. The background contribution
estimated using 
the fit described above is 
$N_b = 45.0 \pm 2.8  \ {\rm (stat.)}$. A parameterisation of
the background model 
shown in figure~\ref{signal}a 
yields a consistent value for $N_b$.
A more
conservative approach is to fit only the power-law background function to  
the full $M (D^* p)$ distribution. The data are compared with the 
result of such a fit in figure~\ref{final}. The corresponding background
estimate is $N_b = 51.7 \pm 2.7 \ {\rm (stat.)}$. 
The probability that a background of $N_b = 51.7$ events fluctuates
to produce at least the number of events in the signal
is $4 \cdot 10^{-8}$, assuming Poisson statistics. 
This probability corresponds to
$5.4 \ \! \sigma$ when expressed as an equivalent number of
Gaussian standard deviations. 
From the change in maximum log-likelihood $\Delta (\ln {\cal L})$
when the full distribution is fitted under the null 
and signal hypotheses, corresponding to the two curves shown in 
figure~\ref{final},
the statistical significance
is estimated to be $\sqrt{-2 \ \Delta (\ln {\cal L})} = 6.2 \ \! \sigma$.

A state decaying strongly to $D^{* \, -} p$ must have
baryon number $+1$ and charm $-1$ and thus
has a minimal constituent quark composition of $uudd \bar{c}$.  
The observed resonance is therefore a candidate for
the charmed analogue $\theta_c^0$ \cite{theory2,cheung} 
of the $\theta^+$. 
The narrow width is reminiscent of that in the strange case.
Given the relatively large mass of the resonance, it is also a 
candidate for an excited state such as the $\theta_c^{* \, 0}$
with spin $3/2$ \cite{close}.

\section{Summary}
\label{concs}

An investigation has been carried out of the invariant mass combinations
of $D^*$ and proton candidates using H1 deep inelastic 
electron-proton scattering data.
A clear and narrow resonance is observed for both
$D^{* \, -} p$ and $D^{* \, +} \bar{p}$ combinations 
with an invariant mass of
$M(D^* p) = 3099 \pm 3 \ {\rm (stat.)} \ \pm 5 \ {\rm (syst.)} \ {\rm MeV}$.
The probability for the background distribution to fluctuate to produce
a signal as large as that observed is less than $4 \cdot 10^{-8}$.
The region of $M(D^* p)$ in which the signal is observed  
contains a richer yield of $D^*$ mesons and exhibits
a harder proton candidate
momentum distribution than is the case for  
side bands in $M(D^* p)$. 
The measured RMS width of the resonance is 
$12 \pm 3 \ {\rm (stat.)} \ {\rm MeV}$, consistent with
the experimental resolution. A signal with compatible mass 
and width is also
observed in an independent photoproduction data sample. 

The resonance is interpreted as an anti-charmed baryon
decaying to $D^{* \, -} p$ and its charge conjugate 
decaying to $D^{* \, +} \bar{p}$.
The minimal constituent quark composition of such a baryon is
$uudd \bar{c}$, making it a candidate for a charmed pentaquark state. 

\section*{Acknowledgments}

We are grateful to the HERA machine group whose outstanding efforts have made
this experiment possible.  We thank the engineers and technicians for their 
work
in constructing and now maintaining the H1 detector, our funding agencies for
financial support, the DESY technical staff for continual assistance and the
DESY directorate for support and for the hospitality which they extend to the
non-DESY members of the collaboration.


\clearpage


\begin{thebibliography}{99}

\bibitem{strangepq}
T.~Nakano {\it et al.}  [LEPS Collaboration],
Phys.\ Rev.\ Lett.\  {\bf 91} (2003) 012002
[hep-ex/0301020]; \\
V.~V.~Barmin {\it et al.}  [DIANA Collaboration],
Phys.\ Atom.\ Nucl.\  {\bf 66} (2003) 1715
[Yad.\ Fiz.\  {\bf 66} (2003) 1763]
[hep-ex/0304040]; \\
S.~Stepanyan {\it et al.}  [CLAS Collaboration],
Phys.\ Rev.\ Lett.\  {\bf 91} (2003) 252001
[hep-ex/0307018]; \\
J.~Barth {\it et al.}  [SAPHIR Collaboration],
Phys.\ Lett.\ {\bf B572} (2003) 127 [hep-ex/0307083]; \\
A.~E.~Asratyan, A.~G.~Dolgolenko and M.~A.~Kubantsev,
submitted to Phys.\ Atom.\ Nucl.\ [hep-ex/0309042]; \\
V.~Kubarovsky {\it et al.}  [CLAS Collaboration],
Phys.\ Rev.\ Lett.\  {\bf 92} (2004) 032001;
erratum-ibid.\  {\bf 92} (2004) 049902
[hep-ex/0311046]; \\
A.~Airapetian {\it et al.}  [HERMES Collaboration],
submitted to Phys.\ Lett.\ {\bf B}\ [hep-ex/0312044]; \\
A.~Aleev {\it et al.} [SVD Collaboration],
submitted to Yad.\ Fiz.\ [hep-ex/0401024]; \\
M.~Abdel-Bary {\it et al.} [COSY-TOF Collaboration],
[hep-ex/0403011].

\bibitem{qcdpq} 
R.~Jaffe, SLAC-PUB-1774, talk presented at the Topical
Conf. on Baryon Resonances, Oxford, UK, 1976; \\
H. H{\o}gaasen and P. Sorba, Nucl.\ Phys.\ B\ {\bf 145} (1978) 119; \\
D. Strottman, Phys. Rev. {\bf D20} (1979) 748; \\
C. Roisnel, Phys. Rev. {\bf D20} (1979) 1646. 

\bibitem{diakonov} 
D.~Diakonov, V.~Petrov and M.~V.~Polyakov,
Z.\ Phys.\ {\bf A359} (1997) 305
[hep-ph/9703373].

\bibitem{na49} 
C.~Alt {\it et al.}  [NA49 Collaboration],
Phys.\ Rev.\ Lett.\ {\bf 92} (2004) 042003 [hep-ex/0310014].

\bibitem{theory2}
R.~L.~Jaffe and F.~Wilczek,
Phys.\ Rev.\ Lett.\  {\bf 91} (2003) 232003
[hep-ph/0307341].

\bibitem{theory1}
H.~Walliser and V.~B.~Kopeliovich,
J.\ Exp.\ Theor.\ Phys.\  {\bf 97} (2003) 433
[Zh.\ Eksp.\ Teor.\ Fiz.\  {\bf 124} (2003) 483]
[hep-ph/0304058]; \\
S.~Capstick, P.~R.~Page and W.~Roberts,
Phys.\ Lett.\ {\bf B570} (2003) 185
[hep-ph/0307019]; \\
M.~Karliner and H.~Lipkin, Phys. Lett. {\bf B575} (2003) 249
[hep-ph/0307243]; \\
D.~E.~Kahana and S.~H.~Kahana,
[hep-ph/0310026]; \\
E.~Shuryak and I.~Zahed,
[hep-ph/0310270].

\bibitem{cspq} H. Lipkin, Phys. Lett. {\bf B195} (1987) 484; \\
C. Cignoux, B. Silvestre-Brac and J. Richard, 
Phys. Lett. {\bf B193} (1987) 323; \\
D. Riska, N. Scoccola, Phys. Lett. {\bf B299} (1993) 338; \\
F.~Stancu,
Phys.\ Rev.\ {\bf D58} (1998) 111501
[hep-ph/9803442].

\bibitem{cheung} 
M.~Karliner and H.~J.~Lipkin,
[hep-ph/0307343]; \\
K.~Cheung,
[hep-ph/0308176].

\bibitem{steinhart} J. Steinhart, `Die Messung des totalen 
$c \bar{c}$-Photoproduktions-Wirkungsquerschnittes durch die
Rekonstruktion von $\Lambda_c$ Baryonen unter
Verwendung der verbesserten $\dedx$ Teilchen\-identifikation am H1 
Experiment bei HERA', Ph.D. thesis, 1999, Universit\"{a}t Hamburg 
(in German,
available from \\
\verb+http://www-h1.desy.de/publications/theses_list.html+).

\bibitem{h1det} H1 Collaboration, 
I. Abt {\it et al.}, Nucl. Inst. Meth. {\bf A386} (1997) 310; \\
H1 Collaboration, I. Abt {\it et al.}, Nucl. Inst. Meth. {\bf A386} 
(1997) 348. 

\bibitem{f2c} 
S.~Aid {\it et al.}  [H1 Collaboration],
Nucl.\ Phys.\ {\bf B472} (1996) 32
[hep-ex/9604005]; \\
C.~Adloff {\it et al.}  [H1 Collaboration],
Nucl.\ Phys.\ {\bf B545} (1999) 21
[hep-ex/9812023]; \\
C.~Adloff {\it et al.}  [H1 Collaboration],
Phys.\ Lett.\ {\bf B528} (2002) 199
[hep-ex/0108039].

\bibitem{pdg} Particle Data Group, H. Hagiwara {\it et al.},
Phys. Rev. {\bf D66} (2002) 010001.

\bibitem{deltam} G. Feldman {\it et al.}, 
Phys. Rev. Lett. {\bf 38} (1977) 1313. 

\bibitem{rapgap} H. Jung, Comp. Phys. Commun. {\bf 86} (1995) 147; \\
H.~Jung, `The RAPGAP Monte Carlo for Deep Inelastic Scattering,
version 2.08', Lund University, 1999, 
(\verb+http://www.desy.de/~jung/rapgap.html+).

\bibitem{lund} B. Andersson, G. Gustafson, G. Ingelman and T. Sj\"ostrand, 
Phys. Rept. {\bf 97} (1983) 31.

\bibitem{jetset} T.~Sj\"ostrand, Comp. Phys. Commun. {\bf 82} (1994) 74 
and LU-TP-95-20 [hep-ph/9508391].

\bibitem{django} 
A.~Kwiatkowski, H.~Spiesberger and H.J.~M\"ohring, Comp. Phys.
Commun. {\bf 69} (1992) 155; \\
L.~L\"onnblad, Comp. Phys. Commun {\bf 71} (1992) 15; \\
K.~Charchula, G.A.~Schuler and H.~Spiesberger, Comp. Phys. Commun {\bf 81} 
(1994) 381, (\verb+http://www.desy.de/~hspiesb/djangoh.html+).

\bibitem{cascade} 
H.~Jung and G.~P.~Salam,
Eur.\ Phys.\ J.\ {\bf C19} (2001) 351
[hep-ph/0012143]; \\
H.~Jung,
Comput.\ Phys.\ Commun.\  {\bf 143} (2002) 100
[hep-ph/0109102].

\bibitem{herwig} G.~Marchesini {\it et al.},
Comp. Phys. Commun. {\bf 67} (1992) 465.

\bibitem{pythia} T. Sj\"{o}strand {\it et al.},
Comp. Phys. Commun. {\bf 135} (2001) 238 [hep-ph/0010017].

\bibitem{belle} 
K.~Abe {\it et al.}  [Belle Collaboration],
submitted to Phys.\ Rev.\ {\bf D}\ [hep-ex/0307021].

\bibitem{jb} A.~Blondel and F.~Jacquet, Proceedings
of the Study for an $ep$ Facility for Europe,
ed. U.~Amaldi, DESY 79-48 (1979) 391.

\bibitem{close} 
J.~J.~Dudek and F.~E.~Close,
Phys.\ Lett.\ {\bf B583} (2004) 278
[hep-ph/0311258]; \\
B.~Wu and B.~Ma, [hep-ph/0402244].
\end{thebibliography}
\end{document}